# Pathway to the PiezoElectronic Transduction Logic Device

**Authors:** P.M. Solomon[1]*, B.A. Bryce[1], M.A. Kuroda[1,2], R. Keech[3], S. Shetty[3], T.M. Shaw[1], M. Copel[1], L-W. Hung[1], A.G. Schrott[1], C. Armstrong[1], M.S. Gordon[1], K.B. Reuter[1], T.N. Theis[1], W. Haensch[1], S.M. Rossnagel[1], H. Miyazoe[1], B.G. Elmegreen[1], X-H. Liu[1], S. Trolier-McKinstry[3], G.J Martyna[1]*, and D.M. Newns[1]*.

**Affiliations:**

[1] IBM T. J. Watson Research Center, Yorktown Heights, New York 10598.
[2] Dept. Physics, Auburn University, Auburn, AL 36849.
[3] Department of Materials Science and Engineering, Pennsylvania State University, University Park, PA 16802.
*Corresponding authors.

TEXT: The information age challenges computer technology to process an exponentially increasing computational load on a limited energy budget[1-3] – a requirement that demands an exponential reduction in energy per operation. In digital logic circuits, the switching energy of present FET devices is intimately connected with the switching voltage[3-5], and can no longer be lowered sufficiently, limiting the ability of current technology to address the challenge. Quantum computing offers a leap forward in capability[6], but a clear advantage requires algorithms presently developed for only a small set of applications. Therefore, a new, general purpose, classical technology based on a different paradigm is needed to meet the ever increasing demand for data processing.

A promising pathway to fast, low voltage classical devices is *transduction* which is widely used in nature to propagate signals in bioorganisms[7]. When propagating digital logic, we require the input and output signal to be electronic – however, this still allows for an intermediate form internal to the logic gates that can induce switching at lower energy than FETs[3-4]. One example of a transductive device is Spintronics (spin transport electronics)[8-9], where the intermediate form is magnetic. Another example is the nanoelectromechanical (NEM) relay[10-11] - a mechanical On/Off switch integrated at the nanoscale - where the intermediate form is the displacement of a cantilever arm.

We have recently proposed a new transduction device, termed the PiezoElectronic Transistor (PET)[12-14], where the intermediate form, *mechanical stress*, is used to trigger an *insulator-metal transition* (IMT) generating the output:

$$\text{input voltage} \xrightarrow{piezo\text{-}transduction} \text{internal stress} \xrightarrow{phase\,change\,(IMT)} \text{large conductance increase} \xrightarrow{electrical\,circuit} \text{voltage output} \quad (1)$$

In Eq. (1) a small input voltage expands a Piezoelectric (PE) pillar, compressing a piezoresistive (PR) element that undergoes a facile IMT, lowering the resistance of the channel and turning the device On – the transition requires little energy to activate. We choose SmSe among possible IMT candidates which include the rare earth monochalcogenides[15] and Mott insulators[16]. A fully integrated PET device, shown in Fig. 1, is predicted by modeling and simulation[12-14] to operate at voltages as low as 0.1V compared with ≈1.0 V for the current technology, the MOSFET[5]. The PET achieves a large power saving while attaining the large On/Off ratio ($\geq 10^4$) necessary for digital logic– an On/Off ratio not available in spintronics[8,9]. The PET can be described as a solid-state relay with all the considerable advantages of NEMS and none of their disadvantages. Indeed, a novel NEM relay[11] has recently achieved an approximately 10 mV switching voltage with an On/Off ratio >$10^{12}$. The PET, however, with its all mechanical parts in continuous contact avoids NEMS stiction and contact-induced wear-out mechanisms which limit cycling[10], and allows for high speed switching, due to its short and stiff mechanical pathway in contrast to the low speed of NEMS. The low resistivity channel in the On-state ensures that the PET can carry large current densities – permitting fast operation due to short RC times. A switch with all of these positive attributes that is manufacturable at high yield, would enable new, fast ultra-lower power computer logic. The concept of piezo based transduction embodied by the PET is a promising pathway to this end.

Interest in this area is growing. Recently another transductive device, also using piezoelectric transducers was proposed[17] where voltage applied to the first transducer causes it to exert a force on the second which outputs an amplified voltage to the gate of an FET.

In this work we present two physical realizations of the PET concept on an early developmental pathway leading to the fully integrated PET of Fig. 1. The two devices are evolved to generate stress and accomplish an IMT in the PR channel - key for demonstrating the viability of the PET concept. The first approach, Gen-0, uses a millimeter-scale piezoelectric actuator to compress a 50 nm thick PR film, metallize the channel and cycle the transition at kHz frequencies. The second, Gen-1, uses a micron scale, lithographed, piezoelectric pillar to compress a nanoscale, e-beam patterned PR element, enabling cycling at 100-kHz frequencies.

The Gen-0 PET generates the stress required to drive an insulator-metal transition in a 50 nm SmSe[18] film where the conducting area is defined by a hole in a silicon nitride layer, as shown in Fig. 2a. A microindenter is utilized as a yoke to provide the counter force against which a commercial piezoelectric actuator compresses and activates conductivity in the SmSe. In operation, a 1 kHz sine wave applied to the actuator with a 20 $V_{p-p}$ (peak-to-peak) amplitude generates a displacement, resulting in a force on the SmSe element. An On/Off modulation of over three orders of magnitude in PR resistance is generated as illustrated in Fig. 2b and c; the same log-linear resistance versus pressure dependence reported earlier under static (DC) conditions[18]. Continued operation of the Gen-0 PET results in stable performance without degradation of the SmSe over $1.25 \times 10^7$ test-time limited cycles (see Fig. 2b). The Gen-0 PET frequency response is bounded by actuator resonance to 1 kHz (note the small phase shift, due to the mechanical delay, between the applied actuator voltage and the PR response), a limitation removed in the Gen-1 device which employs an integrated micro-actuator.

Demonstrating a device with a micro-actuator providing only nanometer sized displacement is key for establishing the viability of the PET concept. The Gen-1 PET, illustrated in Fig. 3a-b, addresses this important challenge. The micro-actuators, fabricated on an 8" silicon wafer, are PE pillars (approximately $2\times2\times1$ μm$^3$), contacted by long leads running on top of patterned PE. Each micro-actuator is flanked by a PE mesh used as a passive support array. The PR elements are fabricated independently on a sapphire wafer (later cut into 2×6 mm$^2$ plates) in a sandwich structure - a bottom metal film, a PR film, and a top metal composed of an array of 400 nm diameter pillars. The full device is assembled by placing the sapphire (receiver) plate on top of the support array and subsequent clamping with the microindenter (Fig. 3a). All electrodes are contacted with pads on the Si wafer. This combination is equivalent functionally to the encapsulated PET of Fig. 1a (see supplementary) with the yoke being the sapphire plate held in place forcefully by the microindenter.

Fabrication of the Gen-1 demonstration PET is complex and challenging; the steps involved are detailed in supplementary material. In brief, the PE actuator pillar and leads are etched into a 1 μm thick blanket PZT ( Pb[$Zr_x Ti_{1-x}$]$O_3$) layer deposited on Pt on the 8" wafer, and capped by a Pt/Ir/Pt composite electrode. The stack was dry-etched using a Ni hard-mask and a hard Ir landing pad was placed on the pillar raising it above the leads. Formation of Al probe pads completed the actuator structure. The PR receiver is fabricated by depositing 50 nm thick SmSe capped by a thin TiN layer on a sapphire wafer, containing a metal pattern. Ti/Ir indenter pillars were formed on this stack by e-beam lithography and lift-off. These pillars were then used as a mask to remove the TiN cap outside the pillars. A micrograph of a PR element is shown in Fig. 3c and a photograph of the full device, including the contacts on actuator side, is shown in Fig. 3d.

Gen-1 PET operation is depicted in Figs. 3a-b: As the microindenter compresses the sapphire plate onto the support array, a single metal pillar element on the plate comes into contact with the landing pad on the PE actuator. The design of the plate and support array allows at least one metal pillar to contact the PE with sub-nanometer vertical precision, thereby controlling the initial stress state of the PR film beneath the metal pillar (see supplementary material). With the microindenter displacement fixed, voltage is now applied across the PE micro-actuator; as the PE expands, it compresses the PR material under the metal pillar reducing its resistance and modulating the current in the output circuit accessed through the bottom metal on the plate.

A static measurement is initiated by slowly lowering the microindenter to compress the sapphire plate while the PR resistance is continuously monitored to detect contact as given in Fig. 3e. Upon contact there is a steep decline in PR resistance, followed by a more gradual decrease as the PR is compressed by the microindenter. The PR resistance vs. microindenter displacement characteristic exhibits hysteresis due to plastic deformation at the plate contacts under high applied microindenter stress. Setting the microindenter displacement to achieve gentle contact, the PE micro-actuator is then activated allowing the switching of the Gen-1 PET shown in Fig. 4a.

In Fig. 4b, the current through the PR is given as a function of voltage across the PR for varying PE voltage, showing the strong modulation of PR current by PE voltage. An On/Off ratio of ≈ 7:1 is achieved. A finite element model (FEM) of the device predicts an On/Off ratio of ≈ 4:1 (see supplementary information). The low observed On-Off ratio (and low transconductance of ~1μS) is attributable to the low mechanical efficiency of voltage transduction to strain in the PR in the Gen-1 design as well as the low-response of the PZT. The fully integrated PET (Fig. 1a)[12-14] is predicted by FEM to have much higher mechanical efficiencies and would use a high response (>10x) piezoelectric material known as PMN-PT[19], which was not available in suitable form for integration during this work. A strong non-linearity of the characteristics is observed (stronger than seen in larger, encapsulated devices[18]). While the cause is unknown, it may easily be attributable to a non-ideal interface on either side of the SmSe.

In the Gen-1 PET, the switch speed of the device is limited by an RC time constant rather than the actuator resonance as in the Gen-0 PET. This RC time is imposed by the need for 1 mm long leads to access the device under the sapphire plate. In this geometry the large parasitic capacitance causes crosstalk between input and output circuits, which increases with frequency. By minimizing the crosstalk electrically, performance of the Gen-1 PET can be observed at frequencies up to 100 kHz (see supplementary information). A Gen-1 PET was subjected to AC cycles of 8 $V_{p-p}$ across the PE, at frequencies of 100 kHz and 50 kHz as depicted in Fig. 4c; only unipolar excursions were applied to minimize PE fatigue. Device operation without degradation was observed for up to $2\times10^9$ cycles. Subsequent failure was due to time dependent dielectric breakdown of the unpassivated PZT[20] rather than in the intrinsic structure.

We have demonstrated the operating principle of the PiezoElectronic Transistor in two forms. These physical embodiments show voltage-to-stress piezoelectronic transduction driving a reversible insulator-metal transition in a PR channel, turning on a solid state switch. Taken together, our results constitute a physical proof of concept for the PET device in demonstrating a significant On/Off switching ratio and resilience under cycling; these devices are a step towards our current build of a fully integrated and later a manufacturable PET. Evolutionary improvements in materials, as discussed above, and fabrication techniques concomitant with a reduction of feature sizes will hopefully enable us to realize the full potential of the PET as a high performance, low power device as predicted by our simulations[12-14]. Apart from the foreseen benefits, our new switching device opens the way for yet unforeseen improvements as the physics of piezo-resistivity and piezo-electricity are investigated on the nanoscale.

FIGURES

**Figure 1.** The PET concept. (a) The fully integrated transductive stack consists of a piezoresistive (PR) element on top of a piezoelectric (PE) element confined by a high Young's modulus (HYM) yoke. The three metal contacts (grey) are termed Gate, Common and Sense. The large PE/PR cross-sectional area ratio of ≈25:1 serves to amplify the stress in the PR relative to that in the PE. (b) Transduction switching principle: The input voltage $V_g$ applied to the gate contact actuates the electrical-to-mechanical transducer (PE) which transmits force to the mechanical-to-electrical transducer (PR). The PR undergoes a facile insulator-metal transition characterized by a log-linear resistivity versus stress response that saturates upon completion of the transition (15). The electrical circuit than generates an output voltage $V_{PR}$ across the PR, gated by the input voltage.

**Figure 2.** The Gen-0 PET. (a) The Gen-0 PET couples a standalone piezoelectric actuator with an integrated solid state PR component, a 50 nm SmSe film deposited in a via structure of 0.5 μm$^2$. A microindenter provides the yoke against which the actuator compresses the SmSe to metallize it. (b) Gen-0 PET actuator voltage (right axis) and PR response (left axis) versus time. The response was stable for 1.25×10$^7$ test-time limited cycles. The actuator voltage is 20 V$_{p-p}$ and the PR is biased at 0.1 V through a high speed logarithmic amplifier which measures current.

(c) Using an FEM model, each point in the PR resistance/actuator displacement time series is converted to a resistance and a pressure in the PR ($p=(1/3)$Tr $T$, $T_1=T_2≈T_3/2$); the results presented in log-linear form have a slope in agreement with static measurements[18].

**Figure 3**. The Gen-1 PET. (a) The device consists of a PE actuator integrated onto a silicon substrate and a metal film, PR film, metal pillar array, sandwich fabricated on a sapphire plate. The PE pillar is 1 μm high and ≈2 μm wide and the metal pillars are ≈0.4 μm in diameter. Upon assembly, the plate is bent under the stress applied by an external microindenter bringing one metal pillar in contact with the metallized top surface of the PE, the "landing pad". (b) Detail of central region of the device showing path of current flow. (c) TEM micrograph cross section of a metal pillar showing the 50 nm thick SmSe layer with a key identifying the layers. (d) Photograph of apparatus. All leads are taken out via electrical contacts on left. The sapphire plate is highlighted in red. (e) Electrical record of contact and compression of the PR element with the indenter, the PE having no bias.

**Figure 4.** The Gen-1 PET in operation. (a) Transfer characteristics of the Gen-1 PET at three values of indenter force. (b) IV characteristics of the PR element at a constant microindenter force with different voltage drops across the PE; in contrast to CMOS transistors, the Gen-1 curves do not saturate - a distinctive feature of transductive devices. (c) Cycling of the Gen-1 PET with PE voltage of 8 $V_{p-p}$ and a constant -4 V offset; the right hand scale indicates number of cycles. The device endured $2\times10^9$ cycles without failure. The failure detected at $7\times10^9$ cycles is due to dielectric breakdown of the PZT under the long leads – a known fatigue mode of these materials if left unpassivated[20] as in this early evolutionary form.

ASSOCIATED CONTENT

Detailed description of device design and fabrication, FEM electro-mechanical simulations, and experimental methods, along with supporting figures. "This material is available free of charge via the Internet at http://pubs.acs.org."

AUTHOR INFORMATION

**Corresponding Author**

Paul M. Solomon, Glenn J. Martyna, and Dennis M. Newns (solomonp@us.ibm.com, martyna@us.ibm.com, dennisn@us.ibm.com).

**Present Addresses**

**Author Contributions**

The manuscript was written through contributions of all authors. All authors have given approval to the final version of the manuscript.

**Funding Sources**

This research was partially supported by the DARPA MESO (Mesodynamic Architectures) Program under contract number N66001–11-C-4109.

**Notes**

ACKNOWLEDGMENT

We wish to acknowledge S. Cordes and E.A. Cartier for help with fabrication and measurement resources, A.S. Kellock for materials characterization, M. Brink for e-beam help, and J.J. Yurkas for technical support. Portions of this work were completed in the IBM T.J. Watson Microelectronics Research Laboratory, Yorktown Heights, NY USA.

REFERENCES

(1) Service, R. F. *Science* **2012**, *335*, 394.

(2) Osborne, I.; Lavine, M.; Coontz, R. *Science* **2012**, *327*, 1595.

(3) Theis T. N.; Solomon P. M. *Science* **2012**, *327*, 1600.

(4) Theis, T.N.; Solomon, P.M. *Proc. IEEE* **2010**, *87*, 2005-2014.

(5) Haensch, W.; Nowak, E.J.; Dennard, R.H.; Solomon, P.M.; *et al. IBM J. Res. Dev.* **2006**, *50*, 339-358.

(6) Ladd, T.D.; Jelezko, F.; Laflamme, R.; Nakamura, Y.; Monroe, C.; O'Brien, J.L. *Nature* **2010**, *464*, 45-53.

(7) Rodbell, M. *Nature* **1980**, *284*, 17–22.

(8) Datta, S.; Das B. *Appl. Phys. Lett.* **1990**, *56*, 665-667.

(9) Awschalom, D.D.; Flatté, M.E. *Nat. Phys.* **2007**, *3*, 153-159.

(10) Liu, T.-J. K.; Alon, E.; Stojanovic, V.; Markovic, D. *IEEE Spectrum* **2012**, *49*, 40-43.

(11) Zaghloul, U.; Piazza, G.; *IEEE Elec. Dev. Lett.* **2014**, *35*, 669.

(12) Newns, D.M.; Elmegreen, B.G.; Liu, X-H; Martyna, G.J. *MRS Bull.* **2012**, *37*, 1071-1076.

(13) Newns, D.M.; Elmegreen, B.G.; Liu, X-H; Martyna, G.J. *Adv. Mat.* **2012**, *24*, 3672-3677.

(14) Newns, D.; Elmegreen, B.; Liu, X-H; Martyna, G.J. *J. Appl. Phys.* **2012**, *111*, 084509.1-18.

(15) Jayaraman, A.; Narayanamurti, V.; Bucher, E.; Maines, R.G. *Phys. Rev. Lett.* **1970**, *25*, 1430-1433.

(16) Z. Yang, C. Ko, S. Ramanathan, *Annu. Rev. Mater. Res.* **2011**, *41*, 337.

(17) Sapan, A.; Yablonovich, E. Nano Lett. **14**, *2012*, 6263−6268.

(18) Copel, M.; Kuroda, M. A.; Gordon, M. S.; Liu, X.-H. *et al*. *Nano Lett.* **2013**, *13*, 4650–4653.

(19) Baek, S.H.; Park, J.; Kim, D.M.; Aksyuk, V.A.; Das R.R. *et al*. *Science* **2011**, *334*, 958-961.

(20) Chen, X.; Kingon, A. I.; Al-Shreef H.; Bellur K. R. *Ferroelectrics* **1994**, *151*, 133-138.

SYNOPSIS: The PiezoElectronic transistor (PET), has been proposed as a transduction device not subject to the voltage limits of field-effect transistors. The PET transduces voltage to stress, activating a facile insulator-metal transition, thereby achieving multi-gigahertz switching speeds, as predicted by modeling - at lower power than the comparable generation field effect transistor (FET). Here, the fabrication and measurement of the first physical PET devices are reported, showing both On/Off switching and cycling. The results demonstrate the realization of a stress-based transduction principle, representing the early steps on a developmental pathway to PET technology with potential to contribute to the IT industry.

TABLE OF CONTENT GRAPHIC;

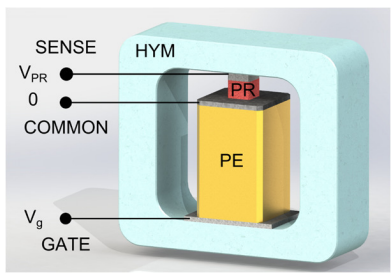

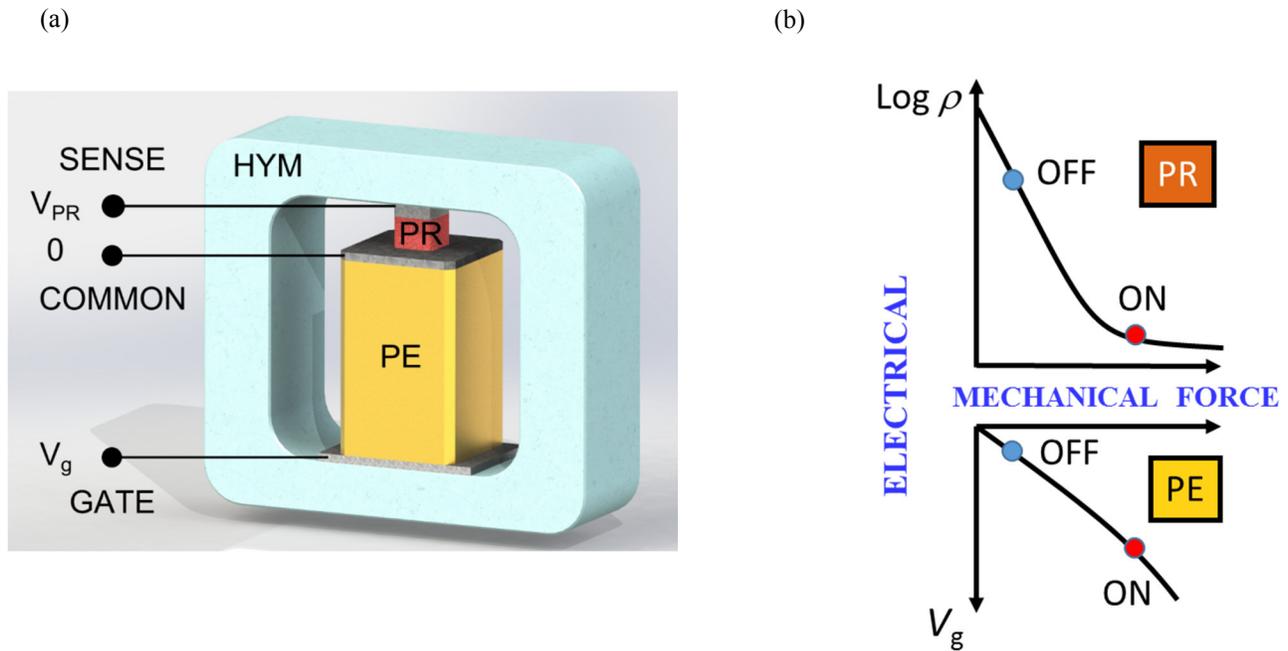

**Figure 1**. The PET concept. **(a)** The fully integrated transductive stack consists of a piezoresistive (PR) element on top of a piezoelectric (PE) element confined by a high Young's modulus (HYM) yoke. The three metal contacts (grey) are termed Gate, Common and Sense. The large PE/PR cross-sectional area ratio of ≈25:1 serves to amplify the stress in the PR relative to that in the PE. **(b)** Transduction switching principle: The input voltage $V_g$ applied to the gate contact actuates the electrical-to-mechanical transducer (PE) which transmits force to the mechanical-to-electrical transducer (PR). The PR undergoes a facile insulator-metal transition characterized by a log-linear resistivity versus stress response that saturates upon completion of the transition (15). The electrical circuit then generates an output voltage $V_{PR}$ across the PR, gated by the input voltage.

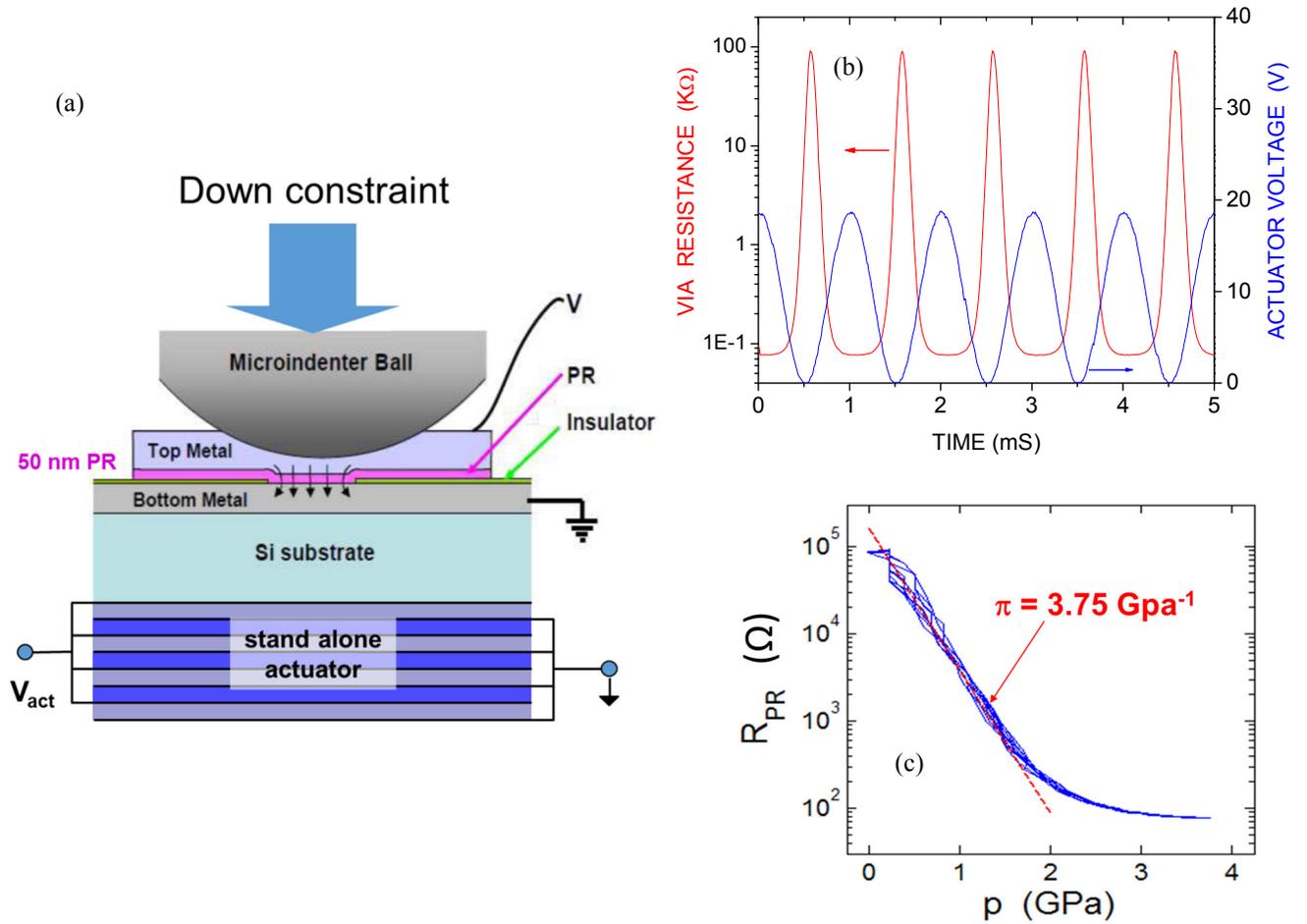

**Figure 2.** The Gen-0 PET. **(a)** The Gen-0 PET couples a standalone piezoelectric actuator with an integrated solid state PR component, a 50 nm SmSe film deposited in a via structure of 0.5 µm$^2$. A microindenter provides the yoke against which the actuator compresses the SmSe to metallize it. **(b)** Gen-0 PET actuator voltage (right axis) and PR response (left axis) versus time. The response was stable for $1.25\times10^7$ test-time limited cycles. The actuator displacement was 150 nm/V and the PR is biased at 0.1 V through a high speed logarithmic amplifier which measures current. **(c)** Using an FEM model, each point in the PR resistance/actuator displacement time series is converted to a resistance and a pressure in the PR (p=(1/3)Tr **T**, $T_1=T_2\approx T_3/2$); the results presented in log-linear form have a slope in agreement with static measurements [18].

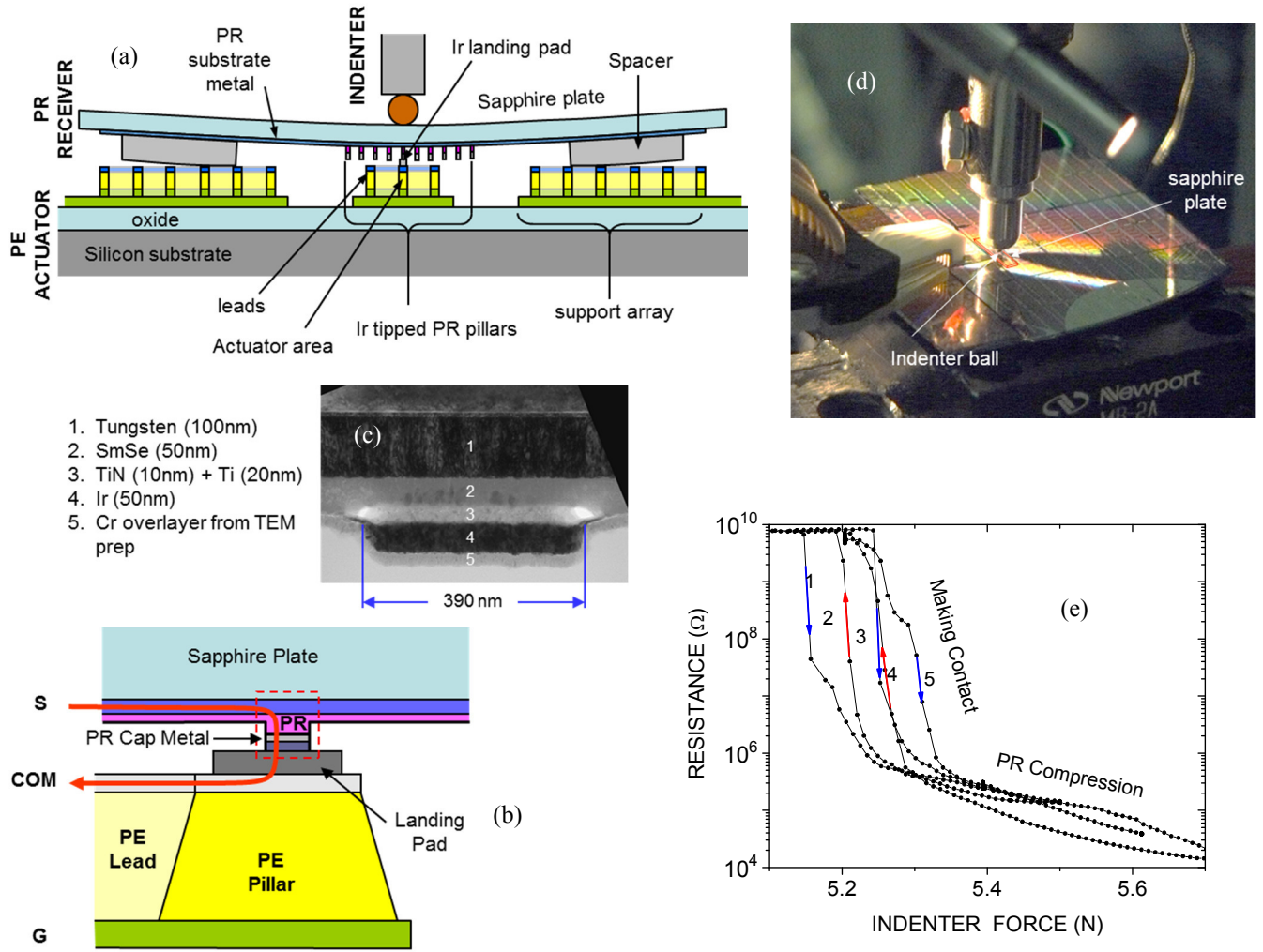

**Figure 3.** The Gen-1 PET. **(a)** The device consists of a PE actuator integrated onto a silicon substrate and a metal film, PR film, metal pillar array, sandwich fabricated on a sapphire plate. The PE pillar is 1 μm high and ≈2 μm wide and the metal pillars are ≈0.4 μm in diameter. Upon assembly, the plate is bent under the stress applied by an external microindenter bringing one metal pillar in contact with the metallized top surface of the PE, the "landing pad". **(b)** Detail of central region of the device showing path of current flow. **(c)** TEM micrograph cross section of a metal pillar showing the 50 nm thick SmSe layer with a key identifying the layers. **(d)** Photograph of apparatus. All leads are taken out via electrical contacts on left. The sapphire plate is highlighted in red. **(e)** Electrical record of contact and compression of the PR element with the indenter, the PE having no bias.

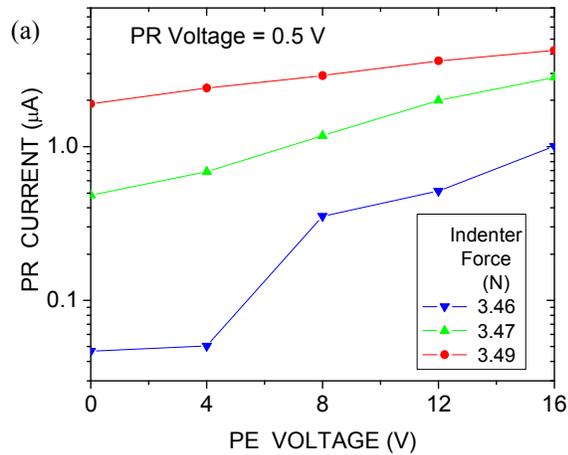

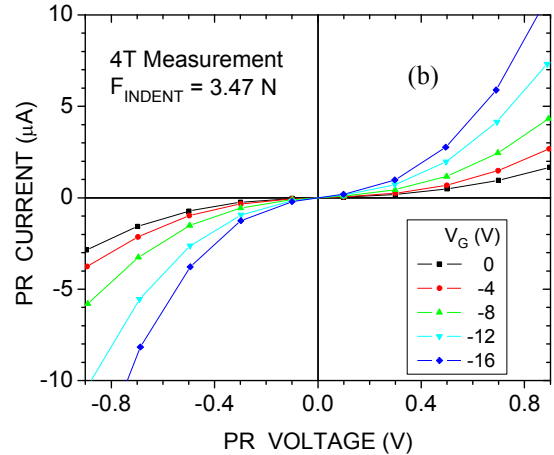

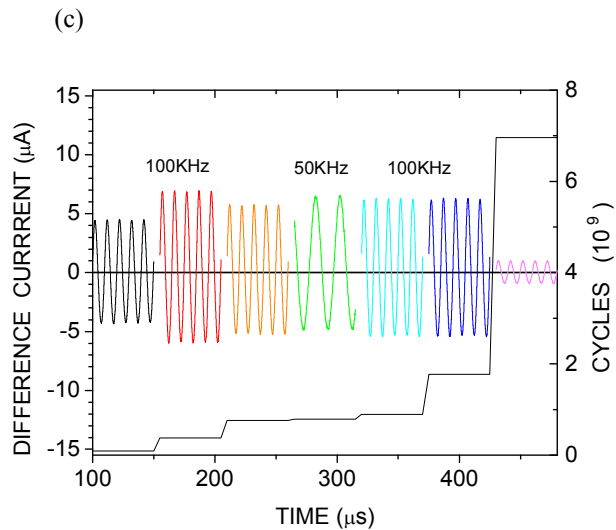

**Figure 4.** The Gen-1 PET in operation. **(a)** Transfer characteristics of the Gen-1 PET at three values of indenter force. **(b)** IV characteristics of the PR element at a constant microindenter force with different voltage drops across the PE; in contrast to CMOS transistors, the Gen-1 curves do not saturate - a distinctive feature of transductive devices. **(c)** Cycling of the Gen-1 PET with PE voltage of 8 $V_{p-p}$ and a constant -4 V offset (the DC current was ~50 μA). The right hand scale indicates number of cycles. The device endured $2\times10^9$ cycles without failure. The failure detected at $7\times10^9$ cycles is due to dielectric breakdown of the PZT under the long leads – a known fatigue mode of these materials if left unpassivated [20] as in this early evolutionary form.

# SUPPLEMENTARY SECTION

## Section 1: PET Characteristics and Circuit Voltage Gain

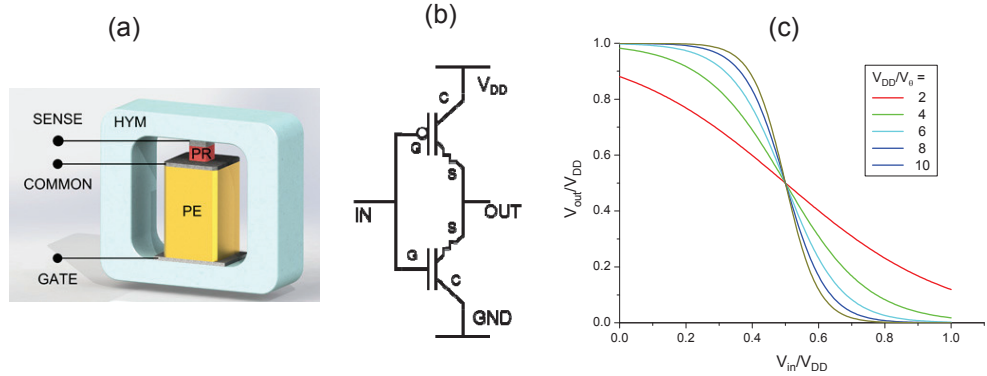

Figure S1-1. (a) PET device identifying the terminals. (b) PET logic inverter circuit including symbols for P (with the circle) and N PETS with the common (C) sense (S) and gate (G) terminals marked. (c) PET inverter transfer curves for various values of $V_{DD}/V_\theta$, where $V_{DD}$ is the power supply voltage and $V_\theta$ the voltage coefficient for the exponential characteristic.

In this section we will briefly recap the PET characteristics and present here the expression for voltage gain derived by Newns et al. [1]. The PET has quasi-exponential transfer characteristic (resistance between sense and common vs. voltage between gate and common) due to the linear transduction of voltage to pressure of the PE and exponential transduction of pressure to resistance of the PR, which can be expressed as:

$$R_{SC} = R_0 \exp(V_{SC}/V_\theta) \tag{S-1}$$

where the more effective the transduction the smaller the coefficient $V_\theta$. In ref 1 a value of 20 mV for $V_\theta$ was quoted. Solving Kirchhoff's equations for the inverter circuit, with no load on the output, results in the equation for the transfer characteristic:

$$\frac{V_{OUT}}{V_{DD}} = \frac{1}{1+\exp\left[\frac{V_{DD}}{V_\theta}\left(2\frac{V_{IN}}{V_{DD}}-1\right)\right]},$$

or alternately as: (S-2)

$$2\frac{V_{OUT}}{V_{DD}} - 1 = \tanh\left[\frac{V_{DD}}{V_\theta}\left(2\frac{V_{IN}}{V_{DD}}-1\right)\right].$$

Equation S-2 is well known since it applies to other established logic families such as bipolar emitter-coupled logic [2].

The transfer characteristic is plotted in Fig. S1-1c for various values of $V_{DD}/V_\theta$. The curves show a robust return to the logic levels ($V_{OUT}=0, V_{DD}$) for $V_{DD}/V_\theta > 6$ i.e. >0.12V in our case.

## Section 2: Mechanical Control of the Split PET

In this section the mechanical design of the split PET is discussed, with an emphasis on achieving sub-nanometer control of displacements along with excellent immunity to outside disturbances as well as excellent tolerance to particulates.

This design replaces the HYM of the PET (see Fig. 1 main text) with a small sapphire plate (SP). The plate is compressed against PZT supports (the support array) by a microindenter. The support array is patterned from the same material stack as the actuator, so they are on the same level. A spacer on the SP raises the PR film surface a specified amount above the actuator so as to clear surface roughness and enable contact to be made with a specified actuator force.

This design has many advantages for an early feasibility demonstration. (1) The structure is simple and does not need any metals to transverse the 1μm thick PZT features in the critical regions of the device. (2) The SP bottom surface is referenced to device-like structures close to the device itself, allowing for a high degree of control of its displacement relative to the device. (3) The open structure of the support array (see Fig. S2-1), along with the 1 μm thick PE, accommodates small particulates which otherwise might cause problems. Small particulates are also crushed by the large force from the microindenter, limiting their effect.

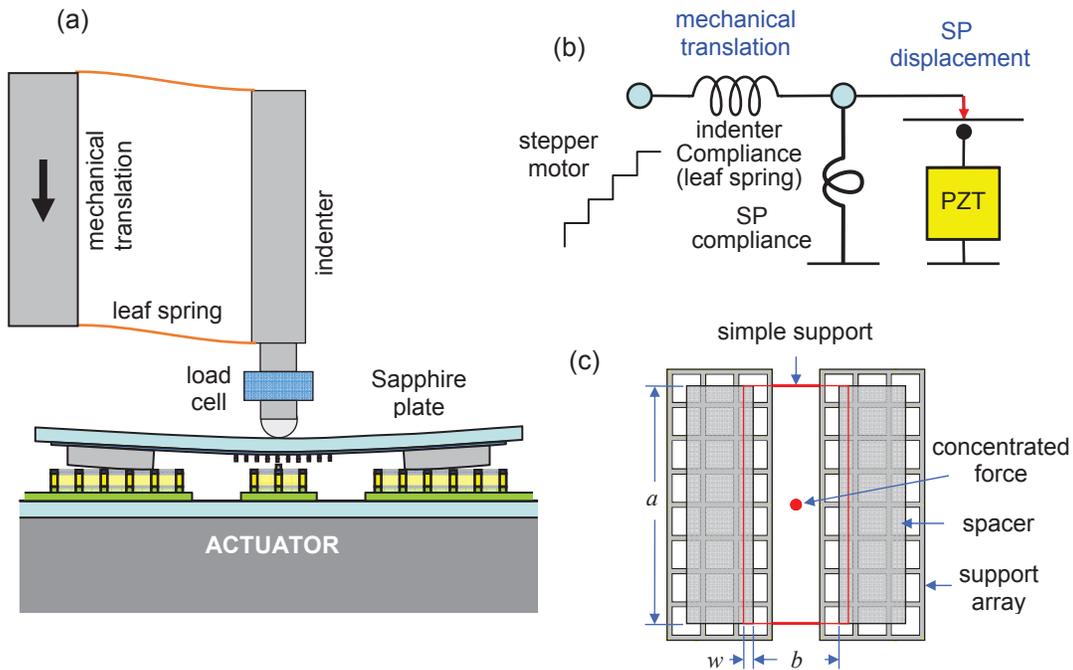

Figure S2-1. (a) Mechanical arrangement for bending the sapphire plate by translating the indenter mount with a stepper motor and generating a controlled force via a compliant leaf-spring. (b) The mechanical equivalent circuit of the arrangement showing that the displacement of the sapphire plate is the displacement of the indenter mount times the ratio of the compliance of the sapphire plate to the compliance of the leaf-spring. (c) Model to derive the compliance of the sapphire plate treating it as a plate with a concentrated central force and simply supported on a rectangular frame.

It was important to verify that the SP is thin enough to flex under the indenter force while simultaneously being rigid enough to serve as an HYM. Firstly, sapphire is a hard material so it would not be indented easily by the local device pressure. Secondly, given a microindenter force $F$ needed to bend the SP through a distance $H$, and given a force $f$ applied via the device from the actuator causing an unwanted SP displacement, $h$, one can show that:

$$\frac{h}{H} = \frac{f}{F}, \quad (S\text{-}3)$$

assuming elastic behavior. Using this formula with typical values, one finds that $f$ is very small due to the small area of the device so that $h$ becomes negligibly small compared to the compression of PR itself, due to $f$. Therefore, the SP acts as a rigid boundary. For instance, for a 200 nm dia. PR at a pressure of 2 GPa, $f$

= $6.3\times10^{-5}$ N, so for $H = 100$ nm and $F = 2$ N we calculate $h$ as .003 nm. This is compared with the compression of the PR of 0.75 nm, assuming a PR thickness of 30 nm and Young's modulus of 77 GPa [3].

The microindenter itself is loosely coupled to its mount via a large compliance leaf spring (see Fig. S2-1a). The vertical translation of the mount, generated with a stepping micrometer, is thus translated into microindenter force in a finely controlled manner. Following the mechanical equivalent circuit of Fig. S2-1b, the mechanical leverage (ratio of the mount translation to the sapphire plate displacement) is equal to the compliance ratio of the leaf spring to the bending of the sapphire plate. This ratio can be made quite large, since the sapphire plate is very stiff and suitably compliant leaf spring can be built.

The large compliance of the leaf spring has the property of mechanically isolating the device from the environment (much like an air table). Once calibrated with the load cell (Fig. S2-1a) the displacement of the mount can be used as a *relative* measure of the indenter force. This procedure is less noisy and more stable than using the load cell directly.

The calibrated compliance of the leaf spring was $C_{LF} = 68$ μm/N. For the SP, the bending compliance was calculated based on a formula for a plate simply supported by a rectangle [4] (Fig. S2-1c). For the case where $b/a > 3$, the compliance, $C_{SP}$, is

$$C_{SP} = \frac{0.185b^2}{Y_{SP}t}, \tag{S-4}$$

where $Y_{SP}$ is Young's modulus of sapphire, $t$ is thickness of the SP and all dimensions are in SI units. For our case, the dimension $b$ is the distance between the two spacers. Because all the force is concentrated on the near edge of the Pt spacers, the spacer will yield, causing the force to be distributed more evenly. The width of the yielded region was estimated to be $w = 26$ μm, using a yield strength for Pt of 200 MPa and a length of $2b$. Once the spacer has been reshaped by yielding, the elastic analysis applies (1.2) but with an effective spacing $b' = b + \frac{2}{3}w$. This gives $C_{SP} = 39$ nm/N. To this we add the compliance $C_{PE}$ of the PZT of 5.8 nm/N assuming a strip of $w$, length $2b$ and area filling fraction of 17%. The compliance ratio $C_{LF}/(C_{SP}+ C_{PE}) = 1500$, so a 1 μm translation of the mount causes a 0.67nm displacement of the SP surface. Because the resolution of the stepping micrometer is 0.05 μm, the SP displacement can be stepped in increments of 33 pm (note that Bohr's radius is 53 pm), at least in a unidirectional manner since the backlash of the micrometer is 2 μm.

Note that an *ad hoc* assumption of length = $2b$ was used for calculating $w$ and $C_{PE}$, rather than the full length $a$. This is to account for weakening of the force in the lateral direction. The choice of this factor is not very critical. For instance, if $b$ rather than $2b$ was used, the compliance ratio would have been 1400.

A test of the capabilities of this system, as well as a way to measure the voltage to displacement transduction factor, $d_{33}$, in the poling direction of the actuator is illustrated in Fig. S2-2. A custom built logarithmic amplifier, based on the LOG114 [5] IC, covering a 0.1 nA to 3 mA range with a frequency response of 10 KHz, supplies a bias of 0.1 V and monitors the current through the device under test (DUT). The DUT can be brought into contact with the actuator either with the stepping micrometer or by applying voltage to the actuator. For this test, the same type of device as in the main text was used but in these measurements contact is barely made, maintaining contact resistances of >100 MΩ. This is to avoid distorting the contact geometry with excessive pressure. The load was first increased at zero actuator voltage to make initial contact; then a voltage sweep was applied to the actuator, keeping the resistance within the 100 MΩ limit. At this point, the indenter mount is raised by 1 μm (i.e. raise the SP by 0.67 nm) and a higher actuator voltage magnitude is applied to bring the device into contact again. This sequence is repeated until the voltage limit of -24 V on the actuator is reached.

The results are shown in Fig. S2-2b and analyzed in Fig. S2-2c, drawing a correspondence between indenter mount translation (hence SP displacement) and the actuator voltage. The LSQ slope of 5.3 V/μm corresponds to an actuator displacement of $10^6/(5.3\times1500) = 126$ pm/V. This is a reasonable value for $d_{33}$ of PZT!

Note that even though the DUT was barely in contact with the actuator, the indenter was firmly in contact with the SP during this measurement with a force of ~1N and this force was barely affected by the small counter-force supplied by the device, most of the reaction coming from the support arrays. Focusing on Fig. S2-2b, an uncertainty (horizontal spread) of about 0.2 µm is reflected in the results. This translates to a control capability of 0.13 nm, justifying our claim in the main text for sub-nm control.

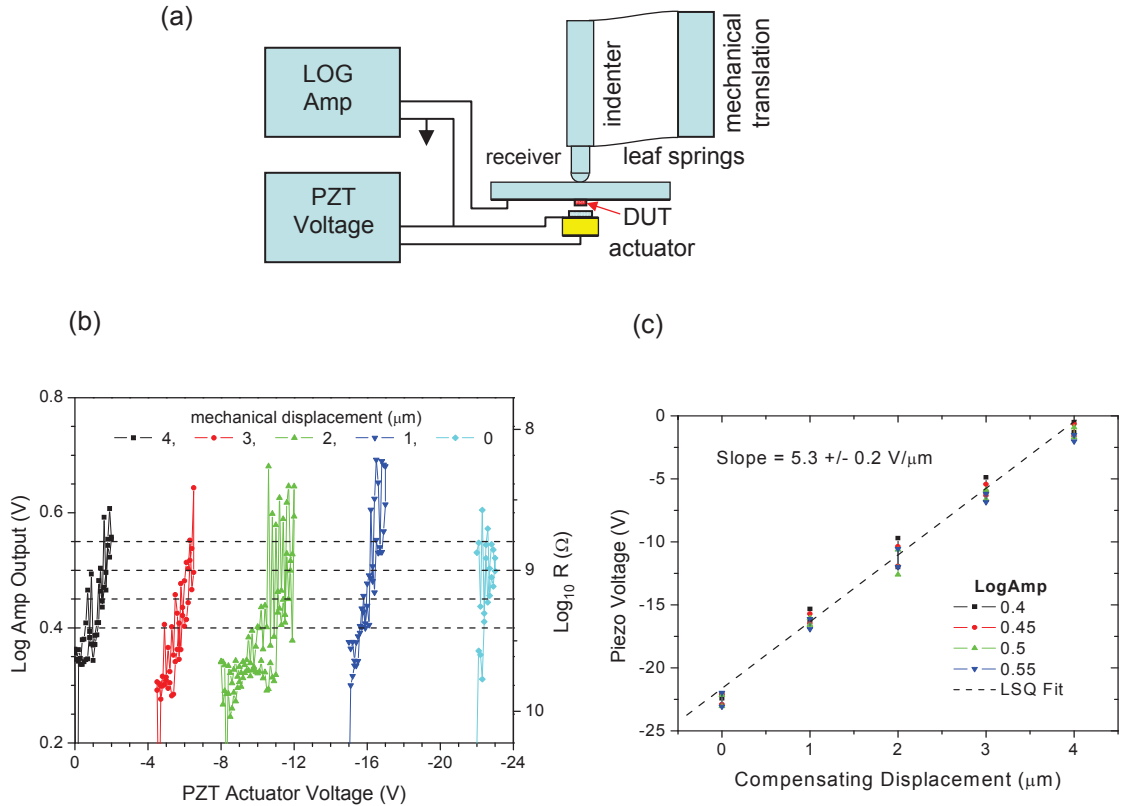

Figure S2-2. (a) Apparatus for measuring the resistance of the contact between the receiver and the actuator as a function of separation. (b) Output of logarithmic amplifier (left) and corresponding log resistance (right), just as contact is made, as a function of the actuator voltage at different displacements of the indenter mount (1µm = 15.7 mN indenter force). (c) Correspondence between indenter mount displacement and actuator voltage needed to restore contact, at a levels of log Amp. Voltage indicated.

## Section 3: Fabrication of Split – PET.

Construction of a piezoelectric transistor (PET) is predicated on patterning a small piezoresistive element and placing that element in intimate contact with the piezoelectric (PE) element to enable the transfer of a large stress to the piezoresistive (PR) element. The materials under study in the present work consist of a $PbZr_{1-x}Ti_xO3$ (PZT) as the piezoelectric (PE) and a SmSe as the PR. These materials present compatibility issues because the former is readily reduced and the latter is readily oxidized. In order to circumvent this situation, as well as to simplify fabrication, a split device was made. The transducer (PE) was fabricated on a silicon wafer, and a receiver (PR) was fabricated on a 100 micron thick deformable sapphire substrate. After independent testing, the two halves were joined into a test device. This removes many processing constraints and has the virtue of creating a test platform with independent control of the PR and PE elements

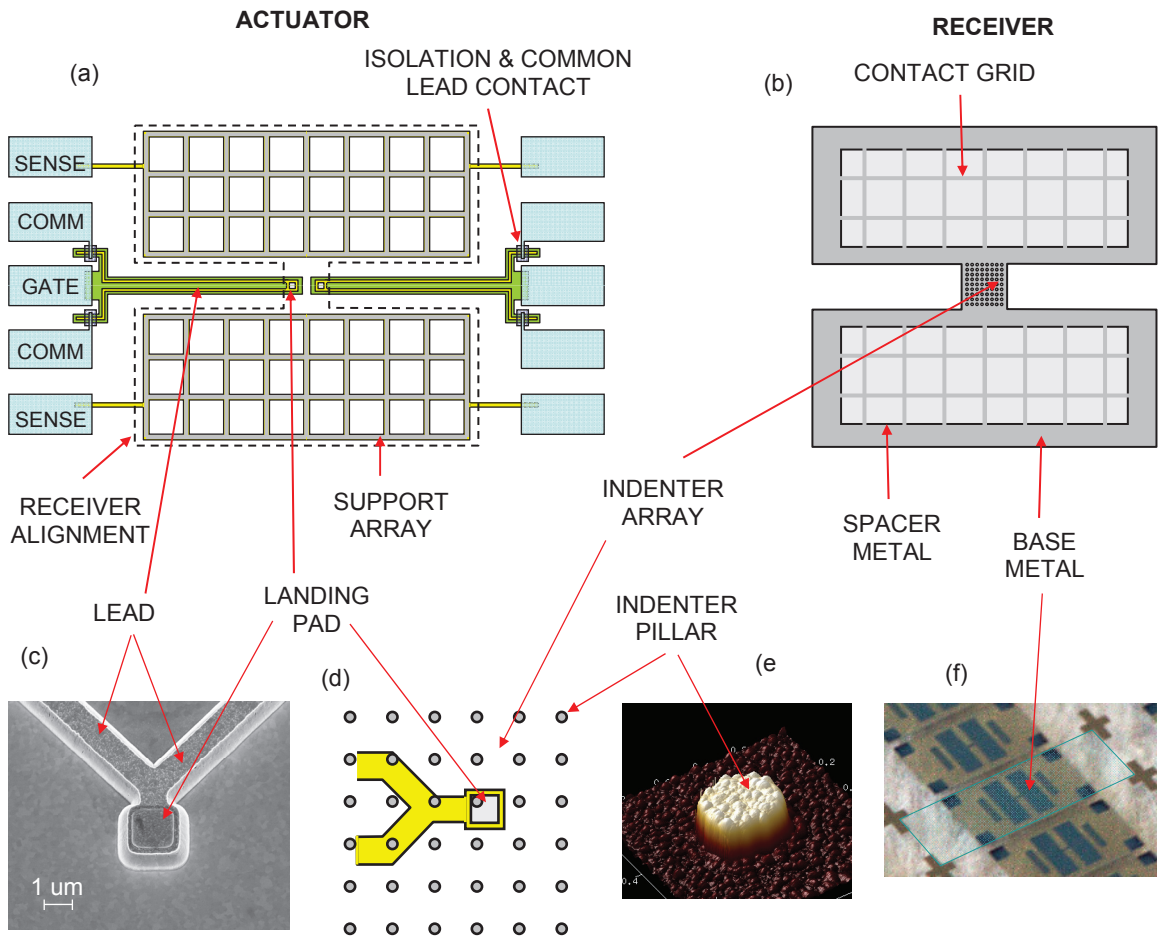

Figure S3-1. Simplified diagrams of actuator containing two devices (a) and receiver (b). Dashed line indicates alignment of receiver on actuator. (c) SEM showing landing pad on top of PZT pillar with 45° angled leads on top of the PZT. (d) Diagram showing random placement of indenter pillar on landing pad. (e) AFM showing detail of pillar. Diameter is ~400 nm. (f) Portion of sapphire wafer containing receiver patterns. Shape of sapphire plate is indicated by highlighted region.

**Design of test structures**

The design of the demonstration PET consists of two parts - an actuator and a receiver section as seen in Fig. 3a of the main text, Fig. 2-1a and Fig. S3-1a&b. The actuator contains a support array which creates a datum at the height of the PE element with the contact area defined by the addition of a metal layer on top of the PE element (the landing pad) in the desired contact region. For experimental convenience all electrical contacts are brought out through the actuator wafer (see Fig. S3-2). The support array provides two electrical paths to the sense terminal via contact with the receiver and the two electrical paths to the common terminal are made through the landing pad. This permits four-terminal measurements which remove the lead resistance in the case of a low-leakage piezoelectric material. The arrangement does not remove any effects of interface resistances between the PR and its metal contacts. The possibility of an oxide hindering the receiver-actuator metal to metal contact is minimized by using Ir and Pt respectively.

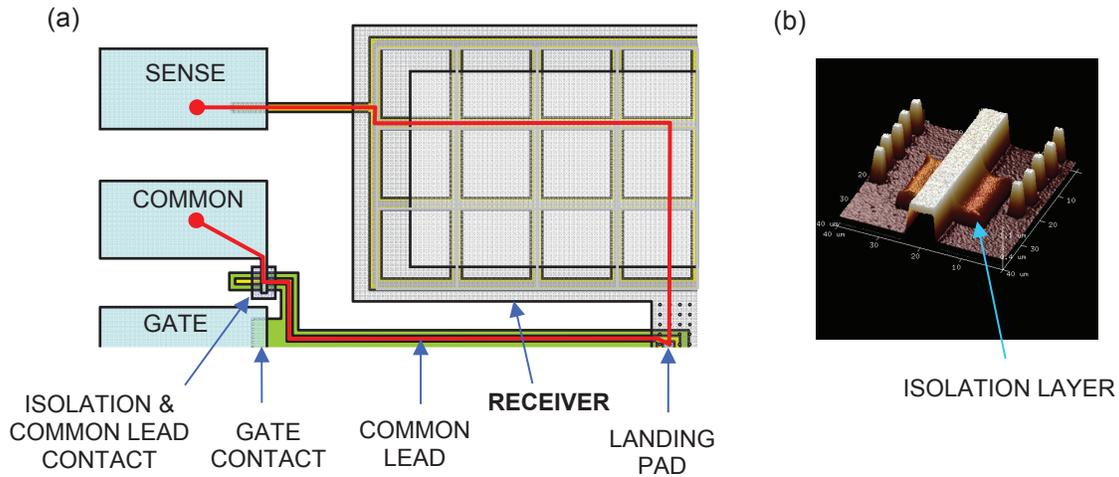

Figure S3-2. (a) Partial diagram (top left quadrant) of the receiver placed on actuator (simplified) showing current path (red line) from COMMON pad on actuator through actuator leads , through indenter contact to landing pad, through metal on receiver and back to actuator and SENSE pad. (b) AFM image showing isolation layer after ashing. The 5 posts on either side are non-functional fill-structures not used for this version of the process.

The receiver section of the device, built on the sapphire wafer, consists of an H shaped structure (Fig. S3-1b). The center bar contains an array of metal dots which transfer stress locally to the PR material when pressure is applied from the PE element (main text Fig. 3b and Fig. S3-1d). The side bars of the H rest upon the support array. The wafer is later diced to create individually mountable sapphire plates (Fig. S3-1f) which may be aligned optically onto the actuator substrate. The sapphire plate is pressed into the actuator using a microindenter. This bends the receiver slightly allowing the metal dots to come into contact with the actuator section of the device. This arrangement also allows for the pressure biasing of the composite structure by varying the force from the microindenter.

**Fabrication of the test structures**

An overview of the fabrication procedure for both receiver and actuator can be seen schematically in Fig. S3-3. To streamline the text flow we use the shorthand $^{\{x\}}$ to reference the sections of this figure, where $x \in [a-k]$.

Fabrication of the actuator section of the PET began with a commercially sourced 1μm thick PZT film. This PZT film was grown on $Pt/TiO_2/SiO_2/<100>$ Si wafer[a]. A layered blanket Pt/Ir/Pt was deposited to thicknesses of 10/30/10nm respectively [b]. Photolithography was performed to enable a liftoff process[c].

This process deposited the raised metal platforms in the contacted region of the actuator section. The metal stack chosen was Ti/Ir deposited to thickness of 10 and 50nm respectively. Liftoff of the metal was performed in N-Methyl-2-pyrrolidone (NMP) with gentle agitation.

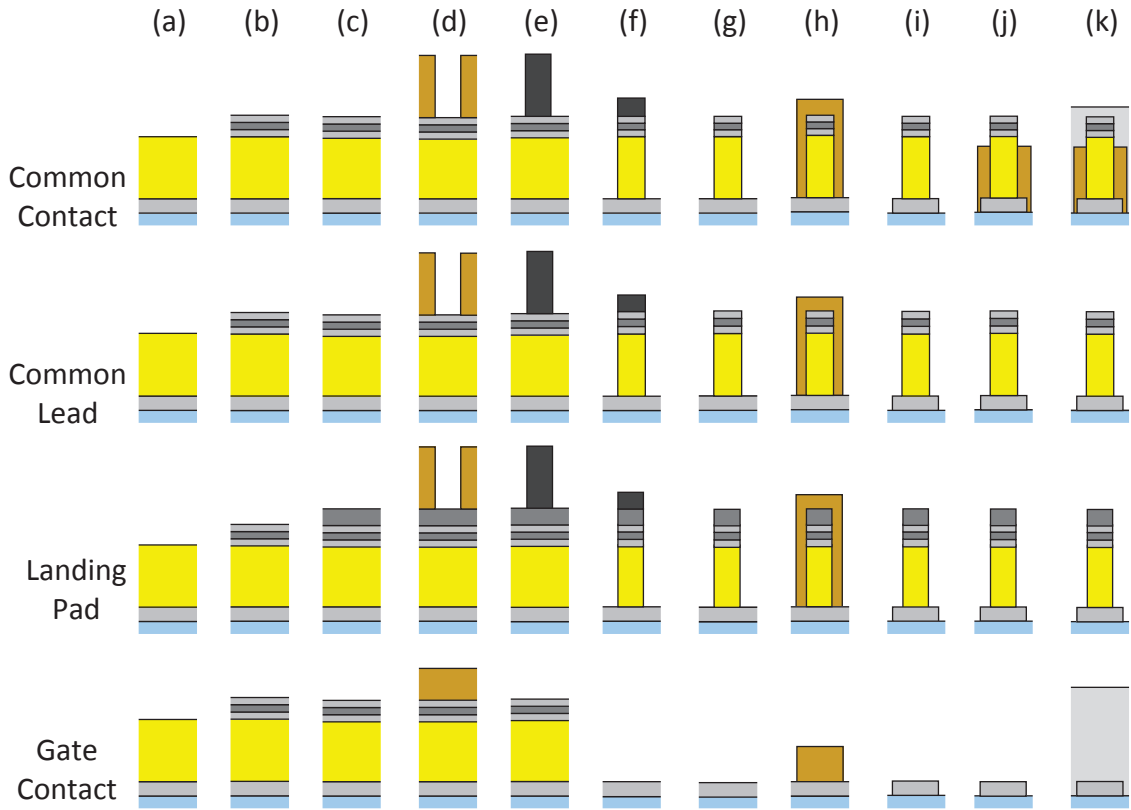

Figure S3-3. Diagrams of the four main functional parts of the split-PET actuator (see Fig. 2-2a) showing their evolution under the eleven main processing steps designated by a-k. The steps are described in the text and referenced by these designations.

Lithography was then used to define the areas to be masked from PE etching (namely the actuator and support structures) {d}. For this purpose, a Ni hard mask was electroplated into the lithographically defined mold. Prior to plating, the resist was ashed in an oxygen plasma to remove surface contamination. Ni was then electroplated {e} into the resist template to a thickness of 400nm. The Ni nucleated on exposed Pt or Ir surfaces. The wafer was then placed in a dual frequency Tegal 6540 RIE chamber and etched [6] cyclically in the plasma using gas flows of 7 sccm $Cl_2$, 28 sccm $CF_4$, and 45 sccm Ar at 300 W (with a 125 W bias power on the wafer) at a chamber pressure of 5 mTorr until the bottom Pt surface below the PZT was reached {f}. The cyclic nature of the etching combined with careful optical observation of the film stack enabled ending the etch at the Pt surface below the PZT. Great care is needed to avoid deposition of resputtered Pt veil on the sidewall of the PE structures which will short the device. Following the PE patterning via RIE, the remaining Ni was removed {g} from the wafer using Transene TFB Ni etch. The etch was heated to 40 °C on a stirred hot plate. The wafer was etched for 2 minutes in this solution and then washed in deionized water.

At this step {g} (SEM in Fig. S3-1c) there are isolated metal capped PE lines on the wafer and a blanket sheet of Pt underneath, covering the entire wafer surface. Patterning of this bottom metal was performed by contact photolithography and ion milling {h} keeping the edge of the bottom metal ~ 4 μm from the PE sidewall (see Fig. S3-2a). Ion milling of the Pt layer leaves Pt residues on the sidewalls of the resist; however these are largely removed when the resist is stripped {i} and therefore have no electrical impact on the final device. Resist is heated and damaged by the milling process and stripping of the modified resist requires a combination of oxygen plasma and hot NMP to completely remove.

To minimize the capacitance and leakage of the contact pad to the PE gate was placed the pad directly on the field oxide. The pad is connected by a metal bridge to the lead on top of the PE. An isolation layer prevents shorting of the bridge to the Pt layer underneath the PE (see Fig. S3-2a&b). A fairly thick layer was needed due to the high gate voltages (~20V) needed. To form this isolation layer, for the simple PET demonstration reported here, photoresist (AZ 1505) was chosen. The photoresist was applied at 5000 RPM and contact lithography performed, with ammonia based image reversal, to create a small bar of polymer over a section of the contact lines to the top of the PE actuator element. This negative tone process was chosen for compatibility with other negative tone polymers more commonly used as stable back end of the line materials. The polymer bar was {j} ashed in a barrel asher until the top metal surface of the PE actuator element was exposed while the sidewalls and bottom metal remained covered. The resist was then baked at 225 C on a proximity hot plate for 30 minutes to reduce the solubility of the resist. This step allowed another pass of contact photolithography to be performed on top the polymer spacer layer to define the metal pads using liftoff {k}. Liftoff of Ti/Al contact pads was performed in NMP with sonication.

The receiver portion (Fig. S3-1b) of the device was formed in a separate process on a 2 inch 100 μm thick sapphire wafer using electron beam lithography for the indenter pillars. To create the receiver 5 nm of Ti and 100 nm of W were sputtered onto the wafer. Contact photolithography was then performed defining a large H shaped structure as well as alignment marks for e-beam lithography. This pattern was transferred into the W and Ti layers using $H_2O_2$ and HF respectively before removing the resist. The next step was to deposit a spacer layer (5 nm of Ti and 125 nm of Pt) on the two arms of the H structure using a shadow mask. The spacer layer's function is to ensure a gap between the receiver and the actuator surfaces so that contact is made only after sufficient pressure is applied.

The PR material, SmSe, and a capping layer, TiN were deposited as described elsewhere [7]. This was performed to a thickness of 30 or 50 nm for the PR layer; the thickness of the capping layer was 10 nm. This capping layer remains on the sample until the sample is ready to be measured to minimize oxidation of the PR material.

Following the deposition of the SmSe and capping layers an array of metal dots and a grid of metal lines are deposited on the receiver using electron beam lithography combined with a PMMA/MMA bi-layer liftoff process. The metal dots are contained in the cross bar of the H-shaped structure and the grids are on the sidebars of the H (see Fig. S3-1b). The grid provides a noble metal contact to the support arrays on the actuator thus completing the electrical contact between the support array and the W counter-electrode on the receiver. The dots, when in contact with the landing pad on the actuator, apply stress to the adjacent PR layer as well as completing the electrical circuit as shown in Fig. S3-2a.

The dots are arranged in an array so that at least 1 dot will contact the landing pad (Fig. S3-1d). For the reported device 10 nm of Ti and 65 nm of Ir were used to create a hard high modulus indenter.

The final fabrication steps are to remove the capping TiN layer and dice the sapphire wafer into individual receiver devices which can be mechanically placed onto the actuator. A wet etch was used to remove the TiN layer selectively to Ti. Following this the receiver wafer was diced into individually mountable 2 mm x 6 mm plates and measured electrically in combination with the already described actuator. A TEM cross-section of final structure is shown in Fig. 3c of the main text.

## Section 4: Finite Element ANSYS™ Simulation of Split PET

The PET performance was characterized by finite element analysis (FEA) using the ANSYS™ software package. Materials were modeled assuming bulk properties. As a first approximation, the piezoelectric coefficients used for the patterned (partially declamped) PZT were those of PZT-4 [8].

The system simulated is described in Fig. S4-1, with Fig. S4-1b showing a detail of the piezoresistive (PR) area contacted by the Ir indenter. For simplicity, a geometry with axial symmetry was assumed. We studied two different cases: the free expansion of the piezoelectric (PE) pillar and the actuation of the piezoresistive (PR) material via the PE material.

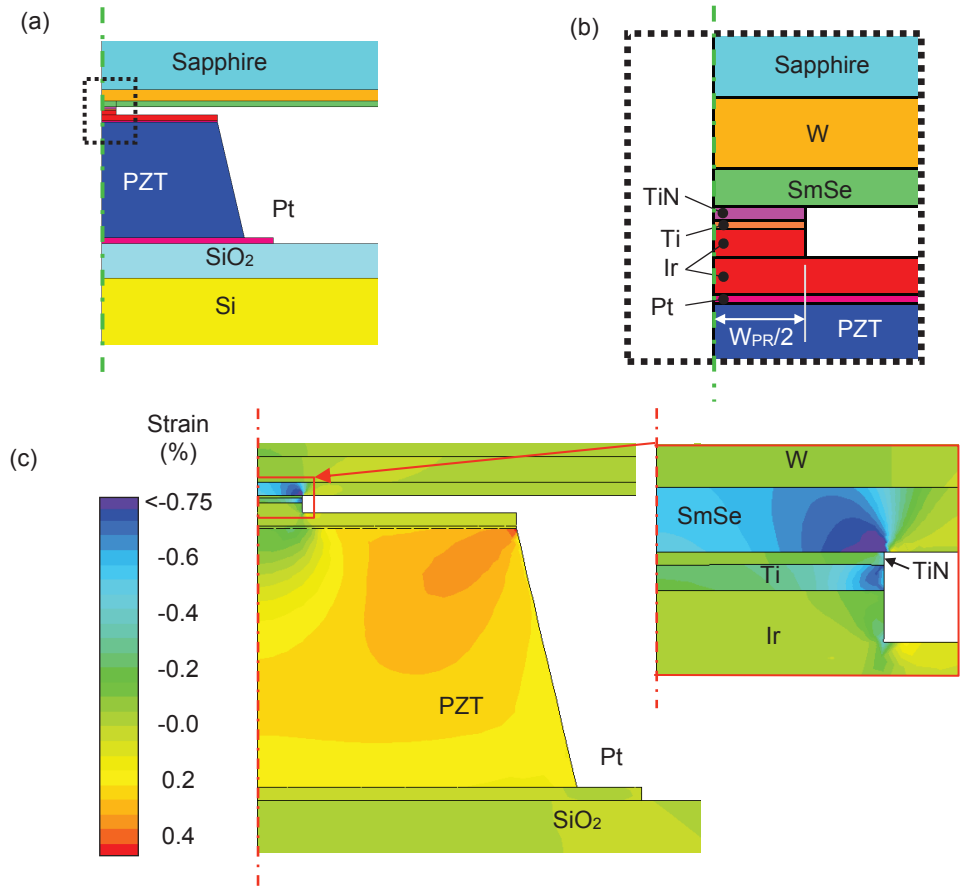

Figure S4-1. (a) Geometry input for ANSYS simulations with detail in (b). The layer thicknesses (bottom to top in nm): Actuator: Si (substrate), SiO$_2$(300), Pt(50), PZT(1000), Pt(10) and Ir(40). Receiver – Ir(60), Ti(20), TiN(10, SmSe(50), W(100) and Sapphire(plate). (c) Strain distribution (perpendicular component) in the region shown in (a) & (b) at 10V across the PE.

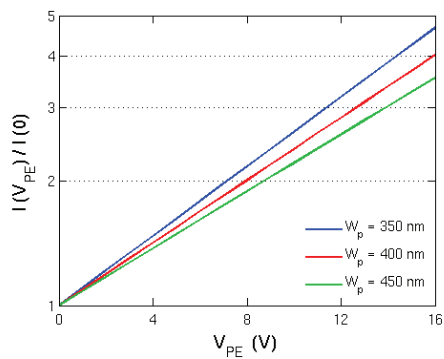

Figure S4-2. On/Off ratio, derived from FEA, plotted as a function of gate voltage and PR width, $W_{PR}$.

In the first case we estimated an effective $d_{33}$ ~ 250pm/V for the displacement of the top of the PE per unit voltage applied to the PZT pillar. This value is larger than the one obtained experimentally (see section 1) but there is a large experimental uncertainty due to the varying contact geometry, suggesting that the PZT-4 coefficients may still be a reasonable description of the PE material. Because of clamping, the effective $d_{33}$ value for this geometry is smaller than the PZT-4 $d_{33}$ bulk value (289 pm/V) expected in systems with larger aspect ratio (height over width).

For the PET device, the stress and strain fields caused by actuation of the PE with an applied voltage were computed. The strain distribution is shown in Fig. S4-1c for an applied voltage of 10V. One clearly sees the overall extension of the PE due to the piezoelectric effect as well as the compression of the PR and corresponding indentation of the PE underneath. To estimate the expected On/Off ratio, we first determine the stress in the PR material, at a given gate voltage, by averaging the hydrostatic stress over the PR region right under the pillar. We next assume that the piezoresistive gauge of SmSe, defined as:

$$\pi_p = -\frac{d \log R}{dp}, \quad (S-5)$$

where $p$ is the pressure, is that observed in thin films [7] which requires about 600 MPa per decade change in resistivity.

Under these assumptions, we expect our system to deliver an On/Off ratio between 3 and 5 at $V_{PE}$ = 16 V (see Fig. S4-2) very similar to the values observed experimentally. Note that because of the small PR thickness compared to $W_{PR}$, most of the current flows in the region with high stress. However, some degradation may stem from the nonuniform stress as well as the spread of the current in regions where the induced stress is low.

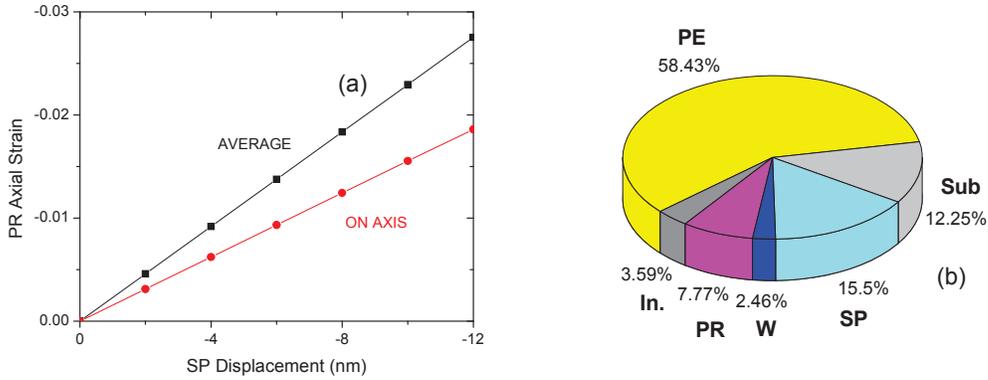

Figure S4-3. (a) Average and on-axis values of the axial component of the PR strain as obtained from FEA. (b) Values of on-axis compression of the different components (sapphire plate, tungsten, PR, Ir indenter and top PE metal, PE and substrate) of the split-PET stack as a percentage of the displacement of the top surface of the sapphire plate.

We use the insights from FEA to further investigate the efficiency of the electro-mechanical transduction of this system. We apply, in the simulation, a relative displacement of the top sapphire plate (SP) and measure the average strain in the PR and also the on-axis displacements of top and bottom surfaces of the sapphire plate, PE and PR (see Fig. S4-3) with the on-axis compression of each component shown in the pie chart. Note that the average compression of the PR will be greater than the on-axis value (Fig. S4-3a) due to the a strong stress concentration near the outer perimeter (Fig. S4-1c). For a total SP displacement of 12 nm, we got an average PR compression of 1.38 nm (about 11%). Examination of the geometry (given in highly exaggerated form in Fig. S4-4) shows where the displacement is being taken up including the desired compression ($\delta_1$) of the PR. Most of the displacement is being lost due to indentation of the PE by the Ir indenter ($\delta_5$) and appreciable amounts due to indentation of the substrate by the PE ($\delta_3$), of the SP by the Ir indenter on the PR ($\delta_4$), and uniform compression of the PE ($\delta_2$). This was supported by rough estimates based on Hertzian indenter theory [9] starting from $\delta_1$ = 1.38 nm and working backwards giving (in nm) $\delta_2$ = 1.23, $\delta_3$ = 0.9, $\delta_4$ = 1.4, and $\delta_5$ = 5.5 with a total of 10.7 nm. The substrate indentation ($\delta_3$) is appreciable in spite of the much lower stress there. This is because the strain is integrated over a larger distance, partially compensating for the lower stress, the other factor being the lower elastic modulus of the substrate $SiO_2$ layer. By indentation theory [9]: $\delta \propto F/(E'D)$, where $F$ is the force (continuous) $E'$ the effective Young's modulus and $D$ the diameter of the region contacted by the indenter.

For this prototype device, we made design choices that aided manufacturability but compromised the On/Off ratio. For instance the Ir layer on top of the PE is not thick enough to adequately act as a rigid anvil as evidenced in Fig. 4.1c by leakage of strain to the PE, and the pie chart. FEA simulations of scaled down, and optimized PET device geometry, using PMN-PT instead of PZT, show that the design goal with On/Off ratios of $10^4$ can be attained.

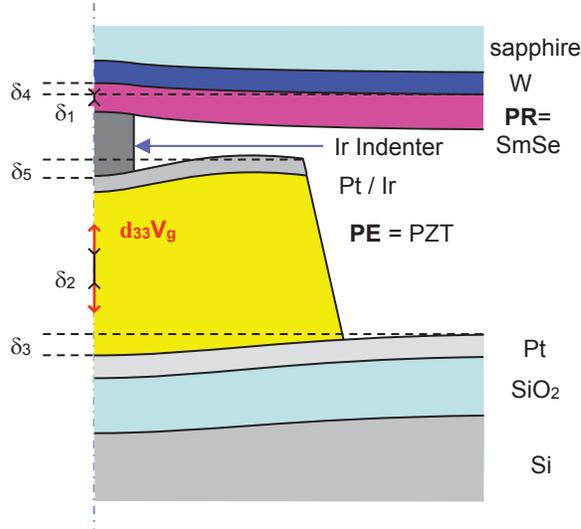

Figure S4-4. Diagram with an exaggerated scale showing the components of displacement, $\delta_1$-$\delta_5$, caused by the PZT actuation with an unloaded top to bottom displacement of $d_{33}V_g$, where $\delta_1$ is the desired compression of the PR, $\delta_2$ is the elastic component of the PE compression, $\delta_3$ is the indentation of the substrate layers by the PE, $\delta_4$ the indentation of the sapphire plate layer by the Ir indenter and $\delta_5$ the indentation of the top PE layers by the same indenter. Typical layer materials are indicated on the right.

## Section 5: Macro PET device for evaluation of Cyclic Switching in SmSe Thin Films

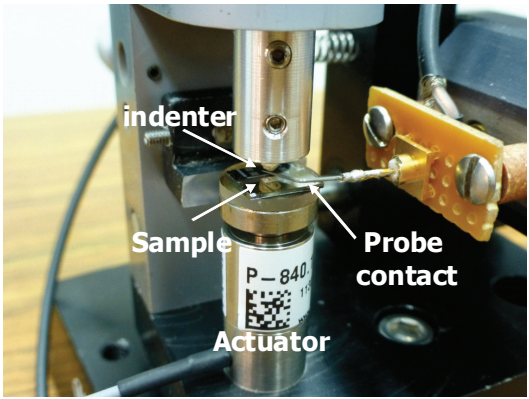
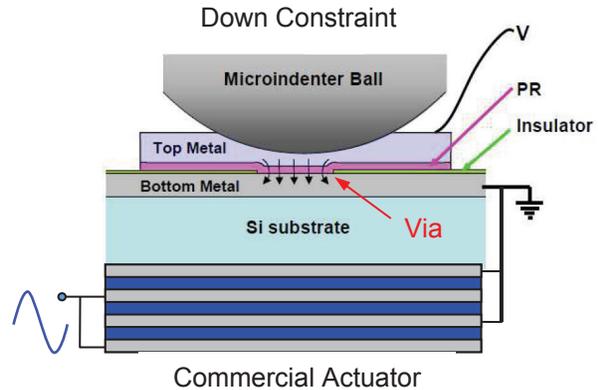

Fig. S5-1 (a) Macro PET device used for cyclic switching of PR films. (b) Schematic showing juxtaposition of commercial actuator, via type PR sample and microindenter.

To evaluate the cycleability of the resistance change that occurs in the SmSe films used for the current device a macroscopic version the PET (Macro PET) was made using a commercial piezoelectric actuator to modulate the stress in the film. The stress was applied to the film using a miniaturized version of the microindenter test system used in previous experiments [7]. A picture of the test fixture is shown in Fig. S4-1a with details of the mechanical arrangement shown in Fig. S5-1b. Reducing the size of the test fixture had two main benefits. Firstly the overall stiffness of the system was increased which enabled higher indentation loads to be achieved with smaller actuator displacements. Secondly the sensitivity to external vibrations was greatly reduced by increasing the overall resonant frequency of the system. The stress applied by the indenter was maximized by using a small diameter tungsten ball (1/64 inch). A pre-stressed piezoelectric actuator that can operate at low loads with frequencies in the kHz frequency range was used to modulate the force applied by the ball. This system incorporates all the essential features a PET device at a macro scale.

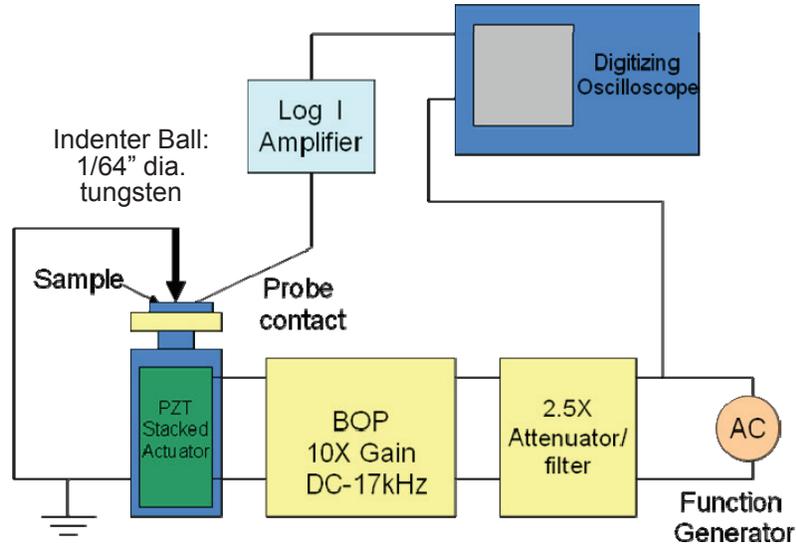

Fig. S5-2 Test circuit for the operation of the Macro PET device. BOP = Bipolar Operational Amplifier (High power positive and negative outputs)

The experimental arrangement for the Macro PET test system is shown in Fig. S5-2. The piezoelectric actuator is driven by a high power bipolar operational amplifier, modulated by input from a function generator. A low pass filter and attenuator was incorporated in the circuit to improve control of the AC input to the actuator. The structure used for testing was the same via type structure used for our previous static loading experiments on SmSe films[7]. The current through the via was monitored using the logarithmic transimpedance amplifier described in section S1. Data from the drive signal on the actuator and the output of the logarithmic amplifier are acquired using a digitizing oscilloscope.

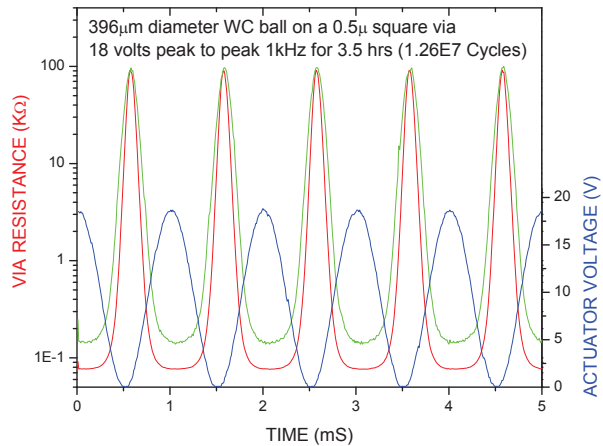

Fig. S5-3 The change in resistance of a 50nm thick SmSe film in a 0.5 µm square via structure under cyclic loading in the macro PET device. The initial resistance response, in green, remains unchanged after greater than $10^7$ cycles (red trace). The voltage applied to the actuator is shown in blue.

Data obtained from a 50nm SmSe film in a 0.5µm square via structure using the Macro PET test fixture are shown in Fig. S5-3. In this experiment a 1 KHz signal with a peak to peak amplitude of 18 V was applied to the actuator. This corresponds to an actuator displacement of approximately 3µm. (Note that only a small fraction of this displacement goes into compressing the SmSe – see below). The vertical position on the indenter was adjusted using a micrometer so that it was continuously in contact with the top electrode of the via structure. Prior to contact the indenter was aligned to the via by visually observing the contact site using a long working distance microscope. The position of the via under that indenter could be finely adjusted by using the electrical probe contact to the sample as a micro manipulator. Optimum positioning of the via under the indenter was obtained by maximizing the electrical response of the sample. The initial response of the via structure measured using and applied voltage of 100mV is shown by the green trace in Fig. S5-3. As can be seen there is a nearly a 1000X change in resistance under the pressure applied by the indenter ball.

The response of the Macro PET via structure was modeled using finite-element analysis (FEA). The simulations showed that the stress/strain fields in the piezoresistive (PR) layer near the indented via are uniform but somewhat anisotropic. It was determined that the nonlinear stiffness of the system (indenting ball/film/substrate) by mapping the force required to make the indenter penetrate the substrate by an amount $\Delta L$. The force (in N for $\Delta L$ in µm) follows:

$$F_S(\Delta L) = A_0 \Delta L^\eta \qquad (S\text{-}6)$$

with $A_0 = 3.18$ and $\eta \sim 1.52$ (close to the 1.5 expected from Hertzian contact mechanics [9]). This force is supplied by the piezoelectric actuator. However as the force is applied the actuator itself is compressed by and equal and opposite force. At 18 volts, the free expansion of the actuator is 2.7 µm. Assuming that the stiffness of the enclosure is ideal, the voltage and displacement in the system can be found by equating the force on the actuator to that on the indenter

$$e_{PE} V_{PE} - k_{PE} \Delta L = A_0 \Delta L^\eta \qquad (S\text{-}7)$$

where the piezoelectric coefficient, $e_{PE} = 8.5$ N/V, and the stiffness of the actuator, $k_{PE} = 57$ N/µm, are obtained from the actuator specifications. The first term calculates the force on actuator under clamped conditions; the second term accounts for the relaxation due to the displacement of the indenter. Solving this relation values for $\Delta L$ are obtained as a function of applied voltage $V_{PE}$. The value of $\Delta L$ includes contributions from deformation of the samples substrate, the top electrode contact as well as the PR film itself. The hydrostatic stress, $p$, in the PR film is obtained from FEA of the via structure at the appropriate value of $\Delta L$ and is used as the independent variable in Fig. S5-4). The piezoresistive gauge (S-3) of 3.75/GPa is obtained from the slope of this plot which is in good agreement with previous static measurements in PR thin films [7].

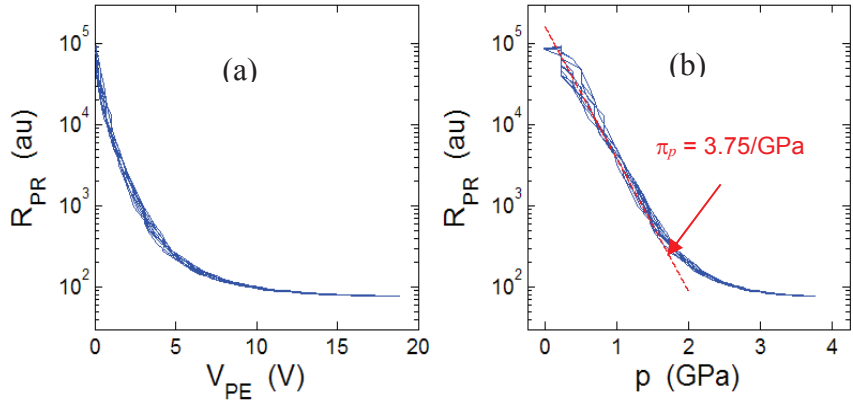

Fig. S5-4 (a) Resistance of the PR film as a function of the voltage in the actuator. (b) Resistance of the PR film as a function of the hydrostatic pressure of the computed using FEA.

A notable feature of the data is the phase shift (9°) between the applied voltage and the resistance response of the sample. Experiments in which the displacement of the sample mount attached to the actuator was monitored using a capacitance gauge indicated that most if not all of this phase shift arises from the dynamic response of the actuator as its resonant frequency is approached. It is this resonance that limits the operation of the Macro PET to a few KHz.

The response of the contact remained approximately constant during the 3.5 hr period the experiment was conducted. A slow drift in the amplitude of the resistive response occurred during testing, but, at all stages of the experiment the original response of the sample could be recovered by a slight adjustment of the vertical position of the indenter. This drift can be attributed to small changes in the vertical position of the indenter due to thermal expansion drift. The response of the sample after 3.5 hrs of operation is shown in the red trace. The resistance modulation has slightly higher amplitude than that observed initially. Small differences in vertical and horizontal positioning of the indenter are thought to be responsible for the observed change. Overall the resistive response of the via after the $1.26\times10^7$ loading cycles of the experiment remains the same. At its lowest resistance, the sample plateaus below 100 Ω. At an applied voltage of 100 mV, this corresponds to 1 mA passing through the 0.5μm via or a current density of $4\times10^5$ A/cm$^2$. The results indicate that the resistance change in SmSe is quite stable at high current densities and is not significantly impacted by cycling to more than $10^7$ cycles.

## Section 6: AC Electrical Tests of Split-PET

To test the endurance of the devices they were subjected to long-duration high-amplitude gate voltages and measured the modulation as a function of time. Pressure was applied to the sapphire plate via the micro-indenter, as described in the main text of this article. The pressure was initially adjusted to achieve a resistance of ~ 100 KΩ in piezo-resistor (PR) element of the device. An AC voltage was then applied to the integrated piezo-actuator (PE) to further modulate the resistance of the PR. The applied voltage was $8V_{p-p}$ with a -4 V offset. This caused a modulation of the current through the piezo-resistor (PR), on the sapphire plate, of about 10%. The waveforms were continuously monitored on a digital oscilloscope.

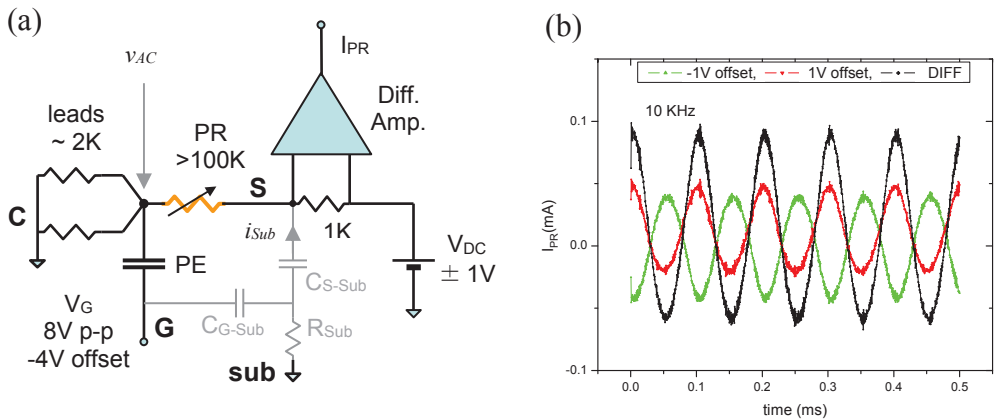

Figure S6-1. AC measurements on Split-PET (a) Measurement circuit, parasitic feedthrough elements shown in grey. The terminals **G**, **C**, **S** and **sub** refer to gate, common, sense and substrate. $C_{G-sub}$ and $C_{S-Sub}$ are capacitances from the bottom PE electrode to substrate due to the gate leads and the support arrays respectively. (b) Measurements at 10 KHz showing change of phase with sign of bias voltage.

An equivalent circuit for the device is shown in Fig. S6-1(a). To clear the sapphire plate the actuator structure is fabricated with long leads (~1 mm) which introduce a resistance in series with the PR of approximately 2 KΩ in each lead. The leads themselves lie on the PZT film with the common electrode on top and the gate electrode on the bottom and so introduce a parasitic capacitive coupling from gate to common. This *rc* combination limits the frequency of operation of the device to <100 KHz. At high frequencies a voltage, $v_{AC}$, is induced on the leads via this coupling which induces a spurious current through the PR, in addition there is coupling through the silicon substrate itself (grayed out elements in Fig. S6-1a) resulting in a spurious current $i_{Sub}$. This coupling occurs (even though the substrate is grounded) via the large capacitances to substrate of the support array, $C_{S\text{-Sub}}$, and the gate leads, $C_{G\text{-sub}}$. The AC current in the PR is given by the equation below:

$$i_{PR} = \frac{1}{R_{PR}}\left(-v_G V_{DC}\frac{\partial \ln R_{PR}}{\partial V_G} + v_{AC}\right) + i_{SUB} \tag{S-8}$$

The first term is the desired current caused by modulation of the PR by the PE due to the AC voltage $v_G$. Note that this term is also proportional to the DC voltage, $V_{DC}$. The second and third terms are due to the spurious $v_{AC}$ and $i_{Sub}$.

A DC voltage, $V_{DC}$, of 1 V is first applied to the PR and the total current through the PR measured by monitoring the voltage across a 1 kΩ resistor connect in series with the PR. The sign of $V_{DC}$ is then reversed and on reversing the pressure modulated current though the PR is inverted (see Fig. S6-1 b) however, the parasitic currents remain the same. Thus, the modulated PR current may be extracted by subtracting the currents at the +/- DC voltages. This subtraction works well except for deviations of the non-linear IV characteristics from odd symmetry (See Fig. 4b of the main text) as can be seen in the results of Fig. S6-1 b for measurements at 10 kHz. As frequency increases $v_{AC}$ increases linearly and $i_{Sub}$ with the square of the frequency limiting this technique to frequencies <100 kHz for our samples.

**References:**
1. Newns, D.; Elmegreen, B.; Liu, X-H; Martyna, G.J. J. A low-voltage high-speed electronic switch based on piezoelectric Transduction. *J. Appl. Phys.* **2012**, *111*, 084509.1-18.
2. Stonham, *T.J. Digital Logic Techniques, Principles and Practices*. Taylor and Francis, US, **1996**.
3. Gupta D.C. and Kulshrestha, S. Pressure induced magnetic, electronic and mechanical properties of SmX (X=Se, Te). *J. Phys.: Condens. Matter,* **2009**, 21 436011 pp10.
4. Sneddon, I.N., The Relation between Load and Penetration in the Axisymmetric Boussinesq Problem for a Punch of Arbitrary Profile., *Int. J. Engng Sci.* **1965**, 3, 47-57.
5. LOG114, Single-Supply, High-Speed, Precision Logarithmic Amplifier, *Texas Instruments*, **2007**, SBOS301A.
6. Griggio, F., Kim, H., Ural, S. O., Jackson, T. N., Choi, K., Tutwiler, R. L. and Trolier-McKinstry, S. Medical Applications of Piezoelectric Microelectromechanical Systems. *Int. Ferroelect*. **2013**, 141 169-186.
7. Copel, M. Kuroda, M. A., Gordon, M. S. & Liu, X.-H. et al. Giant Piezoresistive On/Off Ratios in Rare-Earth Chalcogenide Thin Films Enabling Nanomechanical Switching. Nano Lett., **2013**, 13, 4650–4653.
8. Jaffe, B., Cook, W. R. and Jaffe, H. Piezoelectric Ceramics. *Academic Press*, **1971** reprinted by RAN Publishers, Marietta, OH.
9. Johnson, K. L, Contact mechanics, *Cambridge University Press*, **1987**, Cambridge; New York.